# Analysis of multi-configuration Kohn-Sham methods

Yair Kurzweil and Martin Head-Gordon

*Department of Chemistry, University of California at Berkeley, and,*

*Chemical Sciences Division, Lawrence Berkeley National Laboratory,*

*Berkeley, CA 94720, USA*

We consider the extension of the standard single-determinant Kohn-Sham method to the case of a multiconfiguration trial wavefunction. By applying the *rigorous* Kohn-Sham method to this case, we construct the proper interacting and non-interacting energy functionals. Following the Hohenberg-Kohn theorem for both energy functionals, we derive the corresponding multiconfiguration Kohn-Sham equations. At the end of the analysis we show that, at the ground state, the multiconfiguration wavefunction must collapse into a single-determinant wavefunction, equal to the regular KS wavefunction. We also discuss the non-collapse of the wavefunction in other multiconfiguration density functional theory methods where the auxiliary system is partially interacting.

Density functional theory (DFT)[1] is considered to be the most applicable and useful method to determine the ground-state (GS) physical properties of microscopic electronic systems. Given a system of $N$ particles, bounded in an external potential, $v_{ext}$, the main idea of DFT is searching for their GS *density*, $n$, rather than the full GS WF, $\Psi$. The former can be obtained, in principle, by solving the highly complicated $3N$ dimensional Schrödinger equation (SE). DFT, on the other hand, takes advantage of the Hohenberg-Kohn theorem,[1] which proves that; first, the interacting particles' energy functional, $\varepsilon[\Psi]$, can be replaced by a density functional, $E[n]$

$$E[n] \equiv \varepsilon[\Psi[n]] = F_{HK}[n] + \int dr v_{ext} n,$$
$$F_{HK}[n] \equiv \langle \Psi[n] | \hat{T} + \hat{V}_{ee} | \Psi[n] \rangle, \quad (1)$$
$$\hat{T} = -\sum_{i=1}^{N} \tfrac{1}{2} \nabla_i^2, \quad \hat{V}_{ee} = \frac{1}{2} \sum_{i \neq j=1}^{N} 1/|\mathbf{r}_i - \mathbf{r}_j|,$$

(we use atomic units: $m_e = e = \hbar = 1$) and second, the GS energy and density are obtained from the variational principle of $E[n]$ with respect to the density

$$\delta E[n] \big|_{n=n_{gs}} = 0.$$

However, since the exact one to one map $\Psi \leftrightarrow n$ is unknown, approximations are required in order to estimate the universal HK functional, $F_{HK}[n]$. Therefore Kohn and Sham[2] (KS) suggested the use of an auxiliary trial WF, $\Psi_S$, which partitions $F_{HK}[n]$ as

$$F_{HK} = \langle \Psi_S | \hat{T} | \Psi_S \rangle + E_H[n] + E_{XC}[n], \quad (2)$$

where the Hartree energy, $E_H$, and the exchange-correlation (XC) energy, $E_{XC}$, are defined as

$$E_H[n] = \frac{1}{2} \iint drdr' n(\mathbf{r})n(\mathbf{r}')/|\mathbf{r} - \mathbf{r}'|,$$
$$E_{XC}[n] \equiv F_{HK}[n] - E_H[n] - T_S[n]. \quad (3)$$

It remains to model $E_{XC}$ as an explicit density *functional*, for instance, based on the homogeneous electron gas (the so-called local density approximation (LDA)). The KS auxiliary WF, $\Psi_S$, is taken as a Slater determinant of (fictitious) non-interacting particles, and contains $N$ orbitals, $\{\varphi_i\}_{i=1}^N$. They are determined by the KS method, as is explained later on, such that their density is in principle identical to the exact GS density of the (real) interacting electrons. The density-WF relation is exploited then:

$$n(\mathbf{r}) = N \int dr_2..dr_N |\Psi_S(\mathbf{r}, \mathbf{r}_2..\mathbf{r}_N)|^2 / \langle \Psi_S | \Psi_S \rangle.$$

A present-day grand challenge in DFT is to go beyond the regular KS method and the various widely used XC functionals, *systematically*.[3-9] One direction to improve the regular KS method is searching for the energy minimum using an auxiliary *multideterminantal/multiconfiguration* (MC) trial WF, $\Psi_S^{MC}$, where the energy functional is partitioned similarly to the KS partition, as

$$E = \langle \Psi_S^{MC} | \hat{T} + \hat{V}_{ee} | \Psi_S^{MC} \rangle + E_C^{dyn}[n] + \int dr v_{ext} n, \quad (4)$$

where $E_C^{dyn} \equiv F_{HK}[n] - \langle \Psi_S^{MC}[n] | \hat{T} + \hat{V}_{ee} | \Psi_S^{MC}[n] \rangle$ is to be modeled as an *explicit functional* of the density (it is called the 'dynamical' or 'residual' correlation). Proposed multi-determinantal methods like multi-reference (MR) DFT/CAS-DFT,[3-8] and MC-optimized effective potential (OEP)[9] do exhibit feasibility and improvement over the regular KS method. It is the aim of this letter to examine the rigorous basis of such theories, while extending the *rigorous* regular KS method to the case of a MC trial WF. We show that a *rigorous* MCKS method, i.e. a direct mapping between interacting and *non-interacting* systems, will always yield a single-determinant (SD) WF at the end of the KS-SCF procedure. We also discuss the connection between the above MC methods and our analysis – which clarifies that they generally involve a mapping between interacting and *partially interacting* systems.

To briefly summarize, the KS method[2] introduces an auxiliary system of non-interacting particles, bound in an effective *local* external potential $v_S(\mathbf{r})$, described by the auxiliary wavefunction, $\Psi_S$. A fundamental assumption is that the non-interacting GS density is identical to the interacting GS density, and that this can be restored by a local $v_S$ (v-representability). Since $\Psi_S$ is a Slater determinant of $N$ orbitals $\{\varphi_i\}_{i=1}^{N}$, the energy functional of the *non-interacting* system can be written as

$$E_S[n] = T_S + \int dr v_S(\mathbf{r}) n(\mathbf{r}) \qquad (5)$$

where $T_S = \sum_{i=1}^{N} \langle \varphi_i | -\nabla^2/2 | \varphi_i \rangle$ is the non-interacting kinetic energy functional, the density is given by $n(\mathbf{r}) = \sum_{i=1}^{N} |\varphi_i(\mathbf{r})|^2$, and $v_S$ should be determined. At the same time one can write the *interacting* energy as

$$E[n] = T_S[n] + \int dr v_{ext}(\mathbf{r}) n(\mathbf{r}) + E_H[n] + E_{XC}[n], \qquad (6)$$

We require that the non-interacting and interacting systems have the *same* density at their GS, restored by $v_S$, i.e.

$$\delta E[n_{gs}] = \delta E_S[n_{gs}] = 0. \qquad (7)$$

A crucial part of the KS method is determining the non-interacting potential $v_S$. *The fact that we assume that both interacting and non-interacting systems have the same density at their GS enables us to determine $v_S$ rigorously.* At the GS density both variational principles with respect to the *density* together with the particle number constraint, yield expressions for the constrained variations:

$$\begin{aligned}\delta E[n_{gs}] &= \delta T_S[n_{gs}] + \left(v_{ext} + v_H[n_{gs}] + v_{XC}[n_{gs}] + \mu\right)\delta n \\ \delta E_S[n_{gs}] &= \delta T_S[n_{gs}] + v_S[n_{gs}]\delta n + \mu\delta n\end{aligned} \qquad (8)$$

where $\mu$ is the Lagrange multiplier for the particle number constraint, and,

$$\begin{aligned}v_H[n_0](\mathbf{r}) &= \int dr' n_0(\mathbf{r}')/|\mathbf{r} - \mathbf{r}'|, \\ v_{XC}[n_0](\mathbf{r}) &= \delta E_{XC}[n]/\delta n(\mathbf{r})\big|_{n=n_0}.\end{aligned} \qquad (9)$$

Employing Eq.(8) in Eq. (7), we obtain $v_S$ (up to a constant):

$$v_S[n_{gs}](\mathbf{r}) = v_{ext}(\mathbf{r}) + v_H[n_{gs}](\mathbf{r}) + v_{XC}[n_{gs}](\mathbf{r}). \quad (10)$$

The potential $v_S$ in Eq. (10) defines a map $v_S : E \mapsto E_S$ between the interacting and the non-interacting systems. Furthermore, since $\Psi_S = \det(\{\varphi_i\})/\sqrt{N!}$, the non-interacting energy minimum can also be obtained from the following constrained variation via the orbitals:

$$\delta\left\{E_S + \sum_{i,j}\varepsilon_{ij}\left(\langle\varphi_i|\varphi_j\rangle - \delta_{ij}\right)\right\} = 0, \quad (11)$$
$$\delta E_S \equiv \delta T_S + v_S \delta n.$$

Eq. (11) yields the KS equations[2] for the orbitals:

$$\hat{h}_S(\mathbf{r})\varphi_i(\mathbf{r}) = \varepsilon_i\varphi_i(\mathbf{r}),$$
$$\hat{h}_S(\mathbf{r}) \equiv -\nabla^2/2 + v_S(\mathbf{r}), \quad (12)$$
$$n(\mathbf{r}) = \sum_{i=1}^{N}|\varphi_i|^2.$$

Together with Eq. (10), the KS self-consistent field (SCF) has a closed form. *The KS-SCF ends when $v_S$ yields a set of orbitals that minimizes $E_S$.*

Suppose now that one wishes to extend the KS method by taking the non-interacting trial wavefunction $\Psi_S$ as a MC wavefunction. Thus $\Psi_S^{MC} = \sum_i c_i D_i$, where $\{c_i\}$ are complex amplitudes and $\{D_i\}$ are orthonormal Slater determinants, each composed of a different selection of $N$ orthonormal orbitals chosen from a larger set of $M$ orbitals ($N < M \leq \infty$), and $D_i = \det(\{\varphi_k\})/\sqrt{N!}$. We assume again that there is a local potential, $v_S$, which at the non-interacting GS, generates the GS density of the interacting system. We have to find the GS of the non-interacting energy:

$$E_S^{MC} = \langle\Psi_S^{MC}|\hat{T} + \hat{V}_S|\Psi_S^{MC}\rangle / \langle\Psi_S^{MC}|\Psi_S^{MC}\rangle$$
$$= \sum_{i,j}c_i^*c_j\langle D_i|\hat{H}_S|D_j\rangle / \sum_i|c_i|^2 \quad (13)$$
$$\equiv \sum_{i,j}P_{ij}(\{c_k\})\langle\varphi_i|\hat{h}_S|\varphi_j\rangle,$$

where $P_{ij}$ is the $M \times M$ Hermitian density-matrix and $\hat{H}_S(\mathbf{r}_1..\mathbf{r}_N) = \sum_i \hat{h}_S(\mathbf{r}_i) \equiv \sum_i \left[ -\nabla_i^2/2 + v_S(\mathbf{r}_i) \right]$. Correspondingly, we can define the interacting energy and XC functionals as:

$$E = T_S^{MC} + \left\langle \Psi_S^{MC} \left| \hat{V}_{ee} \right| \Psi_S^{MC} \right\rangle + \int dr v_{ext}(\mathbf{r}) n(\mathbf{r}) + E_C^{dyn}[n],$$
$$T_S^{MC} = \left\langle \Psi_S^{MC} \left| \hat{T} \right| \Psi_S^{MC} \right\rangle, \quad (14)$$
$$E_C^{dyn} \equiv F_{HK}[n] - \left\langle \Psi_S^{MC} \left| \hat{T} + \hat{V}_{ee} \right| \Psi_S^{MC} \right\rangle,$$

where

$$T_S^{MC} = -\frac{1}{2} \sum_{ij} P_{ij}(\{c_k\}) \left\langle \varphi_i \left| \nabla^2 \right| \varphi_j \right\rangle$$
$$n(\mathbf{r}) = \sum_{ij} P_{ij}(\{c_k\}) \varphi_i^*(\mathbf{r}) \varphi_j(\mathbf{r}) \quad (15)$$

Since from Eq. (15) the density is now a functional of the orbitals and the amplitudes $\{c_k\}$, the variation should be done, correspondingly, with respect to each of them. Furthermore, the *same* trial wavefunction $\Psi_S^{MC}$ must be used in the interacting and the noninteracting energy functionals, since the variation is done with respect to the *density*, and the density is a functional of *both* the orbitals and the amplitudes. The objective is again to find a closed form to $v_S$ by equating the constrained variations:

$$\delta E_S^{MC} = \delta E = 0, \quad (16)$$

in analogy to Eq. (7). From Eqs. (14) and (16) we obtain, similarly to the regular KS method:

$$v_S = v_{ext} + v_{XC}^{MC},$$
$$v_{XC}^{MC} = \delta \left\{ \left\langle \Psi_S^{MC}[n] \left| \hat{V}_{ee} \right| \Psi_S^{MC}[n] \right\rangle + E_C^{dyn}[n] \right\} / \delta n. \quad (17)$$

Note that $v_{XC}^{MC}(\mathbf{r})$ can be found similarly to the extended OEP method.[9] The MCKS equations are obtained now from the constrained variation principle

$$\delta E_S^{MC} + \sum_{i,j} \left[ \tilde{\varepsilon}_{ij} \delta \left( \left\langle \varphi_i | \varphi_j \right\rangle - \delta_{ij} \right) \right] = 0, \quad (18)$$

with respect to each orbital and amplitude ($v_S$ is not varied of course). The final result is

$$\sum_j P_{ij}(\{c_k\})\hat{h}_S(\mathbf{r})\varphi_j(\mathbf{r}) = \sum_i \tilde{\varepsilon}_{ij}\varphi_j(\mathbf{r}),$$
$$\frac{\partial E_S^{MC}}{\partial c_k} = 0. \tag{19}$$

Since $\hat{h}_S$ is a 1-body Hamiltonian, orbitals $\{\varphi_i^d\}$ can be chosen to that diagonalize it:

$$\hat{h}_S \varphi_i^d = \varepsilon_i \varphi_i^d, \tag{20}$$

Therefore each determinant, $D^d = \det(\{\varphi_i^d\})/\sqrt{N!}$, is automatically an eigenfunction of $\hat{H}_S$:

$$\hat{H}_S D_i^d = E_i D_i^d,$$
$$E_i = \sum_{j=1}^N \varepsilon_{i_j}. \tag{21}$$

For any such determinant, $E_i \geq E_{min}^d \equiv \sum_{i=1}^N \varepsilon_i$ (assuming that $\{\varepsilon_i\}$ are ordered from lower to higher). We expand the MC-WF in the determinants $\{D_i^d\}$: $\Psi_S^{MC} = \sum_i c_i D_i \equiv \sum_i b_i D_i^d$. Employing the first equality in Eq. (13) and the fact that $\hat{H}_S$ is a sum of single-particle operators (i.e. unitarily invariant), one can see that $E_S^{MC} = \vec{b}^\dagger \ddot{H}^d \vec{b}/|\vec{b}|^2$ where $(\vec{b})_i = b_i$ and $\ddot{H}^d$ is the diagonal matrix $H_{ij} \equiv \langle D_i^d | \hat{H}_S | D_j^d \rangle \delta_{ij} \equiv E_i \delta_{ij}$. Therefore, the non-interacting MC energy is

$$E_S^{MC} = \frac{\sum_i |b_i|^2 E_i}{\sum_i |b_i|^2} = E_0 + \frac{\sum_i |b_i|^2 \Delta E_i}{\sum_i |b_i|^2} \tag{22}$$

where $E_0 = \min\{E_i | b_i \neq 0\}$ and $\Delta E_i = E_i - E_0$. Thus, provided that $\Psi_S^{MC}$ contains more than one nonzero $b_i$, and $E_0 = \sum_{j=1}^N \varepsilon_{i_j} \geq E_{min}^d$ as we concluded above, we obtain:

$$E_{min}^d \leq E_0 < E_S^{MC}. \tag{23}$$

Therefore, for *any* given local potential, $v_S$, the MC-WF will never minimize $E_S^{MC}$ unless it collapses into a single determinant (In the case of degeneracy, the MC-WF will collapse into a sum of determinants containing non-excited orbitals). We conclude that Eqs. (16),(17) can never be satisfied rigorously by a nontrivial MC-WF.

Despite the fact that the rigorous MC-KS simply collapses to the standard KS method, there are MC-DFT methods that do not exhibit such a WF collapse. For instance, the CAS-DFT/MR-DFT approach[3-5, 7, 8] uses the partition (4) to model the total energy by a MC-WF. The energy functional is then mapped onto a 'partially' interacting system via an effective local potential, $w$, $w : E \mapsto \bar{E}$, where

$$E = \left\langle \Psi_S^{MC} \middle| \hat{T} + \hat{V}_{ee} \middle| \Psi_S^{MC} \right\rangle + E_C^{dyn}[n] + \int dr v_{ext}(\mathbf{r}) n(\mathbf{r})$$
$$\bar{E} \equiv \left\langle \Psi_S^{MC} \middle| \hat{T} + \hat{V}_{ee} \middle| \Psi_S^{MC} \right\rangle + \int dr w(\mathbf{r}) n(\mathbf{r})$$

The assumption here is that there is a local potential, $w$, that at the GS of the *partially interacting* $\bar{E}$, generates the same *fully interacting* GS density of $E$. $w$ is determined from the HK variational principle (with total particle number constraint): $\delta E = \delta \bar{E} = 0$ to yield $w(\mathbf{r}) = v_{ext}(\mathbf{r}) + \delta E_C^{dyn} / \delta n(\mathbf{r})$. This equality defines a consistent map between the interacting and non-interacting systems, if the variation of $\bar{E}$ is done with respect to both orbitals and amplitudes. Hence the amplitudes and the orbitals are determined by the SCF equations for the amplitudes, $\partial \bar{E} / \partial c_i = 0$, and for the orbitals

$$\delta \left[ \bar{E} + \sum_{j,k} \varepsilon_{jk} \left( \left\langle \varphi_j \middle| \varphi_k \right\rangle - \delta_{jk} \right) \right] / \delta \varphi_i = 0,$$ This variation leads to extended KS equations, containing both local and *nonlocal* potentials (see Eq. 20 in Ref. [8]). WF collapse is not likely here, since the KS equations contain nonlocal potential, reflecting partial treatment of electron-electron interactions in $\left\langle \Psi_S^{MC} \middle| \hat{T} + \hat{V}_{ee} \middle| \Psi_S^{MC} \right\rangle$. Therefore, $\bar{E}$ cannot be written as a *bilinear* form of the amplitudes as was done for $E_S$ in Eq. (13).

Another method is the MC-OEP,[9] a variation of CAS-DFT, with a difference regarding the evaluation of the orbitals. In this method, the orbitals are obtained from the regular KS equations (12), where $v_S$ is the local OEP, calculated from the variation $\delta \bar{E} / \delta v_S = 0$. Here, again, MC-WF collapse is unlikely, since the amplitudes are calculated for the partially interacting system, and then, with frozen values of the amplitudes, only the orbitals have to minimize a non-interacting energy. Therefore MC effects enter the KS equations only implicitly, through $v_S$. There is no global stationary principle, however.

In summary, we proved that there is no direct map between interacting and non-interacting systems, if the latter is defined by a nontrivial auxiliary wavefunction that is multi-configurational. On the other hand, a map between interacting and partially interacting systems, the latter described by MC-WF, does not lead to a WF collapse to the single determinant KS WF.

Acknowledgement: This work was supported by the Director, Office of Energy Research, Office of Basic Energy Sciences, Chemical Sciences Division of the U.S. Department of Energy under Contract DE-AC0376SF00098.